\documentclass[12pt]{article}

\usepackage{graphicx}
\usepackage{color}
\usepackage[colorlinks=true, urlcolor=blue, linkcolor=blue, 
citecolor=blue]{hyperref}
\usepackage{cite}

\def \lsim{\mathrel{\vcenter
     {\hbox{$<$}\nointerlineskip\hbox{$\sim$}}}}

\newcommand{\beq}{\begin{equation}}
\newcommand{\eeq}{\end{equation}}
\newcommand{\beqa}{\begin{eqnarray}}
\newcommand{\eeqa}{\end{eqnarray}}
\newcommand{\beqar}{\begin{eqnarray*}}
\newcommand{\eeqar}{\end{eqnarray*}}

\begin{tiny}

\end{tiny} 

\begin{document}
\thispagestyle{empty}
$\,$

\vspace{32pt}

\begin{center}

\textbf{\Large Origin of the high-energy neutrino flux\\[1ex]  
at IceCube}

\vspace{50pt}
J.M.~Carceller$^a$, J.I.~Illana$^a$, M.~Masip$^a$, D.~Meloni$^b$
\vspace{16pt}

\textit{$^a$CAFPE and Departamento de F{\'\i}sica Te\'orica y del Cosmos}\\
\textit{Universidad de Granada, E-18071 Granada, Spain}\\
\vspace{7pt}
\textit{$^b$Dipartimento di Matematica e Fisica, 
Universit\`a di Roma Tre}\\
\textit{Via della Vasca Navale 84, 00146 Rome, Italy}\\
\vspace{16pt}

\texttt{jmcarcell@correo.ugr.es, 
jillana@ugr.es, masip@ugr.es, meloni@fis.uniroma3.it}

\end{center}

\vspace{30pt}

\date{\today}

\begin{abstract}
We discuss the spectrum of the different components in the astrophysical
neutrino flux reaching the Earth and the possible contribution of each component
to the high-energy IceCube data. We show that
the diffuse flux from cosmic ray interactions with gas
in our galaxy implies just 2 events among the 54 event sample. 
We argue that the neutrino flux from cosmic ray 
interactions in the intergalactic (intracluster) space
depends critically on the transport parameter $\delta$ 
describing the energy dependence in the 
diffusion coefficient of galactic cosmic rays. 
Our analysis motivates a $E^{-2.1}$ neutrino spectrum with a
drop at PeV energies that fits well the data, including the 
non-observation of the Glashow resonance at 6.3 PeV. 
We also show that a cosmic ray flux described by an unbroken
power law may produce a neutrino flux with interesting 
spectral features (bumps and breaks) related to changes
in the cosmic ray composition.
\end{abstract}

\newpage

\section{Introduction}
The IceCube observatory has recently discovered a flux of
TeV--PeV neutrinos whose origin is {\it not} atmospheric
\cite{Aartsen:2013jdh,Aartsen:2013jdhb,Aartsen:2015zva}. 
In particular, four years of data 
include a total of 54 events of energy above 28 TeV.
The spectrum, the track to shower ratio and the angular
distribution of these events are consistent with a diffuse
(isotropic) flux of astrophysical neutrinos harder than
the atmospheric one. Where do these neutrinos come from?

Cosmic rays (CRs) are the key to understand any (atmospheric, 
galactic or extragalactic) neutrino fluxes, and
it seems clear that IceCube's discovery will have 
profound implications in CR physics. Moreover, 
the remarkable simplicity observed in the 
CR spectrum suggests that the high-energy neutrino flux may
admit an equally simple description. Let us be more specific.

We observe that at energies below $E_{\rm knee}=10^{6.5}$ GeV 
CRs reaching the Earth are dominated 
by hydrogen and $^4$He nuclei, having both species slightly 
different spectral index \cite{Boezio:2012rr,Yoon:2011aa}. 
In particular, we estimate
\beq
\Phi_p = 1.3 \left( {E\over {\rm GeV}} \right)^{-2.7}
\; {\rm particles / (GeV\,cm^2\,s\,sr)}
\label{fluxp}
\eeq
and 
\beq
\Phi_{\rm He} = 0.54 \left( {E\over {\rm GeV}} \right)^{-2.6}
\; {\rm particles / (GeV\,cm^2\,s\,sr)}\,.
\label{fluxHe}
\eeq
These expressions imply an all-nucleon flux 
$\Phi_N \approx 1.8 \,( {E/ {\rm GeV}} \,)^{-2.7}
\;{\rm (GeV\,cm^2\,s\,sr)^{-1}}$ and a similar number of
protons and He nuclei at 
$E\approx 10$ TeV.
Beyond $E_{\rm knee}$ up to 
$E_{\rm ankle}=10^{9.5}\; {\rm GeV}$ the CR
composition is uncertain, while the total flux becomes
\beq
\Phi = 330 \left( {E\over {\rm GeV}} \right)^{-3.0}
\; {\rm particles / (GeV\,cm^2\,s\,sr)}\,.
\label{fluxA}
\eeq
These spectral features, together with the almost perfect
isotropy observed in the flux 
and the primary to secondary CR composition 
(the B/C ratio, 
the frequency of antimatter or of radioactive nuclei in CRs) can be 
accommodated within the following general scheme. 

Galactic CRs are accelerated according to a power law
$E^{-\alpha_0}$. The spectral index $\alpha$ that we
see would then result after including propagation effects:
CRs {\it diffuse} from the sources and 
stay trapped by galactic magnetic fields \cite{Han:2006ci}
during a time proportional 
to $E^{-\delta}$. As a consequence 
$\alpha=\alpha_0+\delta$, expressing that higher energy CRs
are less frequent both because they are produced at a lower 
rate and because they propagate 
with a larger diffusion coefficient and 
leave our galaxy faster. 
The transport parameter
$\delta$ is universal, in the sense that it is identical
for CRs with the same rigidity
$R=E/(Ze)$, and its value would be determined by
the magnetic fields in the 
interstellar (IS) medium. For example, a Kraichnan or
a Kolmogorov spectrum of magnetic turbulences imply
$\delta=1/2$ or $1/3$, respectively 
\cite{Ptuskin:2006xz}. The value
of $\alpha_0$, in turn, may include some dependence with 
the CR composition; notice, in particular, that the difference
in the proton and He spectral indices observed at 
$E<E_{\rm knee}$ requires
that $\alpha_0^{p} \approx \alpha_0^{\rm He}+0.1$. As for
the spectral break observed 
at $E_{\rm knee}$, it is thought to be associated 
to the sources rather than the transport: up to subleading
effects  \cite{Blasi:2012yr}
$\delta$ could be constant at all energies between
10 GeV (where the effects of the heliosphere become important)
and a critical energy near $E_{\rm ankle}$ where the Larmor radius
of CRs equals the maximum scale of the magnetic turbulences
in the IS medium.

Although the basic parameters $\alpha_0$ and $\delta$ 
can be fit using the observables
mentioned above, there is some degree of degeneracy that allows
for different possibilities \cite{Maurin:2010zp,DiBernardo:2009ku}. 
Take the CR flux  below $E_{\rm knee}$ in 
Eqs.~(\ref{fluxp},\ref{fluxHe}). 
The spectral index $\alpha=2.6$--$2.7$ in the proton and
He fluxes may result from 
$\alpha_0=2.1$--$2.2$,
which is expected from diffusive acceleration at
supernova remnants, with $\delta=0.5$. The data, however, could
also be fit with a smaller diffusion parameter $\delta=0.3$--$0.4$
if $\alpha_0\approx 2.3$, which could be explained in models
with significant CR reacceleration, or even
with $\delta\approx 0.8$ in scenarios with strong convective winds
if $\alpha_0\approx 1.9$.

Here we will argue that the analysis of the 
high-energy IceCube signal gives us not only an indication of its
origin, but also it may provide a hint of what 
the value of $\delta$ is. Our objective is 
to discuss the spectrum of the different components in the diffuse 
neutrino flux reaching the Earth and propose one of these components 
as the main source of the IceCube neutrinos. In particular,  we will show that 
the simplest case with $\delta=0.5$ implies a consistent picture where
the bulk of the signal comes from neutrinos produced 
in the extragalactic (intracluster) medium.

First we will review the expected contribution to the IceCube data set
from 
neutrinos produced in CR interactions inside our own galaxy. 
This will let us estimate the excess of events
relative to the atmospheric plus galactic background. Then we will discuss 
the spectrum of two extragalactic components in the diffuse $\nu$ flux,
and we will calculate their possible contribution to the
IceCube signal. Finally, we will consider the possibility that 
the flux discovered by IceCube is not
a power law: we will show that the interactions 
of a CR flux described by a single power law may introduce 
bumps and breaks in the secondary neutrino spectrum
associated to sudden changes in the CR composition at different energies.

\section{Diffuse flux of galactic neutrinos}
Astrophysical neutrinos of $E>1$ GeV are non thermal, they 
appear always as secondary particles 
produced in the collisions of high-energy CRs 
with matter. Let us start considering the ones produced
inside our own galaxy. CR collisions may take place 
mainly in two different environments: the interstellar (IS) medium 
where CRs are trapped for a long time (diffuse flux) and 
the dense regions at or near the acceleration sites (pointlike sources). 
The diffuse $\nu$ flux comes predominantly from 
directions along the galactic disk, 
whereas the local sources include pulsars and supernova remnants.

The diffuse flux of galactic neutrinos has been estimated by
a number of authors \cite{Stecker:1978ah,Berezinsky:1992wr,
Strong:1998fr,Evoli:2007iy,Joshi:2013aua,Ahlers:2013xia}. 
It basically depends on three factors:
{\it (i)} The CR density at each point in our galaxy, {\it (ii)} 
the gas density in the disk and the halo, and {\it (iii)} 
the neutrino yield in the collisions of the CRs with the IS
gas (hydrogen and helium in a proportion near 3 to 1 in mass). 
We will take the approximate analytical expressions for 
the diffuse flux obtained
in \cite{Carceller:2016upo}, where we can find a 
detailed account of these three factors. At 
$10^{3}$--$10^{5.5}\,{\rm GeV}$ this neutrino flux is
[in ${\rm (GeV\,cm^2\,sr\,s)^{-1}}$]
\beq
\bar \Phi_\nu^{\rm gal} = 
3.7\times 10^{-6} \left({E\over {\rm GeV}}\right)^{-2.617}
+
0.9\times 10^{-6} \left({E\over {\rm GeV}}\right)^{-2.538}
\,,
\label{fluxnuG0}
\eeq
where the two terms correspond to the contributions 
from the protons and the He nuclei in the CR flux, respectively, 
and the uncertainty is estimated at the $20\%$.
Eq.~(\ref{fluxnuG0}) provides the total neutrino plus antineutrino
flux averaged over all directions. The angular dependence
($\Phi_\nu$ is 100 times stronger from the galactic disk than from high 
latitudes) can also be found in \cite{Carceller:2016upo}.
After oscillations the relative frequency of each flavor reads
\beq
(\nu_e : \nu_\mu : \nu_\tau : \bar \nu_e : \bar \nu_\mu : \bar \nu_\tau )
=  ( 1.13 : 1.07 : 0.99 : 0.91 : 0.99 : 0.91 )\,,
\label{flav1}
\eeq
for the component in the flux coming from protons and 
\beq
(\nu_e : \nu_\mu : \nu_\tau : \bar \nu_e : \bar \nu_\mu : \bar \nu_\tau )
= ( 1.04 : 1.06 : 0.97 : 1.00 : 1.00 : 0.93 )\,. 
\label{flav2}
\eeq
for the neutrinos from He (or from any nucleus with a similar 
number of protons and neutrons).

At energies $E\ge 10^{5.5}\,{\rm GeV}$
the expression in Eq.~(\ref{fluxnuG0}) is
no longer valid, and at $E> 10^{6.5}\,{\rm GeV}$ the flux
$\bar \Phi_\nu$ is determined by the 
the CR composition  beyond $E_{\rm knee}$ \cite{Carceller:2016upo}:
\beq
\bar \Phi_\nu^{\rm gal} = \left\{\begin{array}{ll}
4.4 \times 10^{-4} \, \left( {E\over {\rm GeV}}\right)^{-2.918} 
& {\rm \; (100\%\;proton)}\,,\\ 
1.2 \times 10^{-4} \, \left( {E\over {\rm GeV}}\right)^{-2.938} 
& {\rm \; (100\%\;helium)}\,,\\ 
1.3 \times 10^{-5} \, \left( {E\over {\rm GeV}}\right)^{-2.974} 
& {\rm \; (100\%\;iron).}\,
\end{array}\right.
\label{fluxnuG1}
\eeq
We will use these expressions and will interpolate with a
power law in the $10^{5.5}$--$10^{6.5}$ GeV energy interval.

\begin{figure}[!t]
\begin{center}
\includegraphics[scale=0.8]{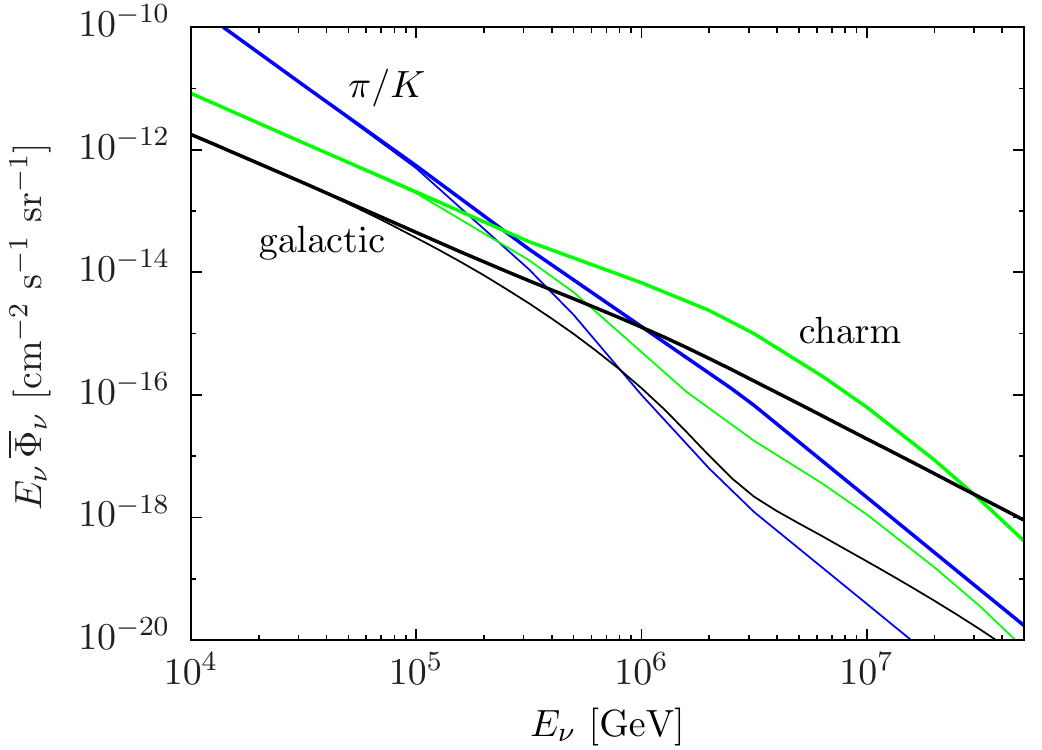}\hfill
\includegraphics[scale=0.8]{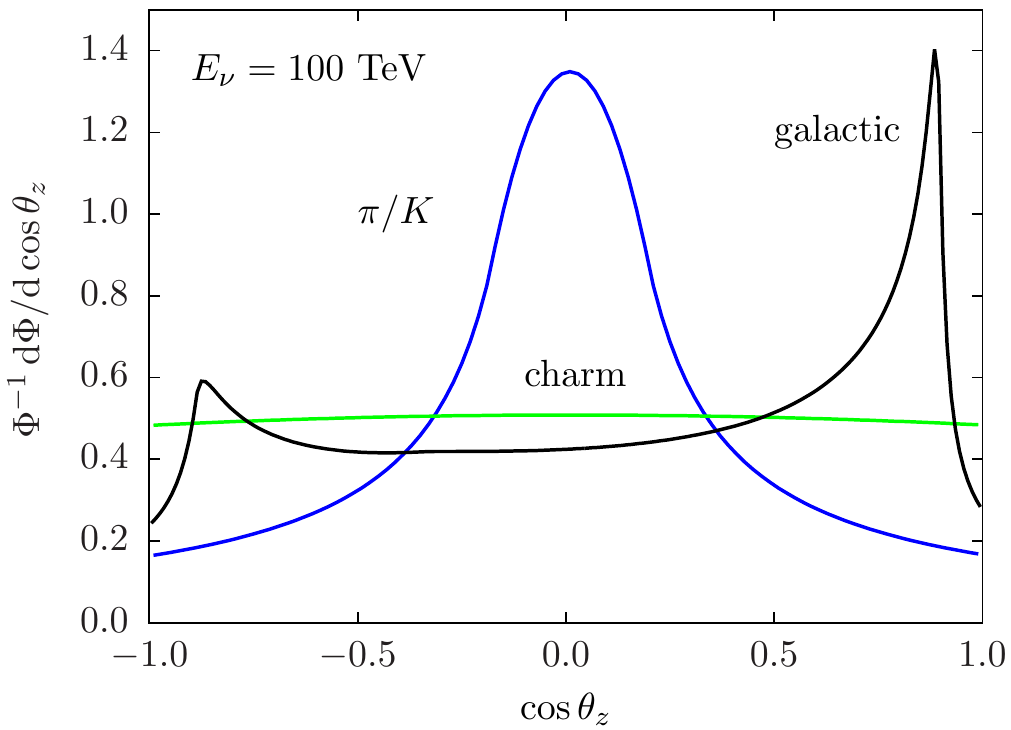}
\end{center}
\caption{{\bf Left.} Galactic and atmospheric fluxes (conventional
and from charm decays); the thick (thin) lines correspond to
a 100\% proton (iron) CR composition at $E>E_{\rm knee}$. 
{\bf Right.} Normalized 
zenith angle distribution of 
these neutrino fluxes at IceCube ($\delta=\theta-\pi/2$).
\label{f1}}
\end{figure}
If compared with the atmospheric $\nu$ flux,
the galactic flux in Eqs.~(\ref{fluxnuG0},\ref{fluxnuG1})
is small but not negligible. 
The atmospheric flux has two main 
components: the so called conventional neutrinos from light-meson
decays (a detailed calculation can be 
found in \cite{Lipari:1993hd}) and  neutrinos from the 
prompt decay of forward charm \cite{Halzen:2016thi}. 
In Fig.~\ref{f1} we plot these fluxes and their dependence
with the zenith angle $\theta$ at IceCube, where the 
declination is just $\theta-\pi/2$. At these energies 
the conventional flux contains muon and electron neutrinos in an
approximate 30 to 1
proportion, whereas the $\nu$ flux from charm 
decays has a similar frequency of both flavors  
and a $2\%$ of $\nu_\tau$. Conventional and charm neutrinos
are described by different spectral indices (see Fig.~\ref{f1}),
and although the first ones dominate the atmospheric flux up to 
$E\approx 250$ TeV,  charm hadrons are the main source of 
electron neutrinos already at 10 TeV \cite{Halzen:2016thi}.
The atmospheric $\nu$ flux also has a 
strong dependence on the CR composition at $E>E_{\rm knee}$ (it 
is proportional to $1/A$ \cite{Lipari:2013taa,Gaisser:2013ira}).
In the plot we see that 
the diffuse galactic flux is below the conventional
one at $E<1$ PeV, and it is just one fourth of the flux 
from charm decays at all IceCube energies.

We can readily estimate the number of events that these fluxes 
imply at IceCube and compare it with the data. In Table~\ref{tab1} we have
defined three energy bins, three direction bins, and we have
separated shower from track events \cite{Illana:2014bda}.
\begin{table}[!t]
\begin{center}
\begin{tabular}{r|c|c|c||c|c|c||c|c|c|c}
          \multicolumn{1}{c}{}
        & \multicolumn{1}{c}{Data} 
        & \multicolumn{1}{c}{$\;$Atm$\;$} 
        & \multicolumn{1}{c}{$\;$Gal$\;$} 
        & \multicolumn{1}{c}{Data} 
        & \multicolumn{1}{c}{$\;$Atm$\;$} 
        & \multicolumn{1}{c}{$\;$Gal$\;$} 
        & \multicolumn{1}{c}{Data} 
        & \multicolumn{1}{c}{$\;$Atm$\;$} 
        & \multicolumn{1}{c}{$\;$Gal$\;$} 
& \\
\multicolumn{11}{c}{}\\ [-3ex]
\cline{2-10} 
{\scriptsize Tracks } & 2 & 0.6+0.2 & 0.0 & 1 & 0.1+0.1 & 0.0 & 0 
& 0.0+0.0  & 0.0 &  
{\scriptsize UPGOING }\\ [0.5ex]
\cline{2-10}
{\scriptsize Showers}& 6 & 0.2+1.0  & 0.2  & 1 & 0.0+0.5 & 0.1 &0& 
0.0+0.1 &0.0& 
{\tiny ($+20^\circ\!\!<\delta<\!\!+90^\circ$) }\\
\cline{2-10}
\multicolumn{11}{c}{}\\[-2ex]
\cline{2-10}
{\scriptsize Tracks } & 3 & (3)+2.8+0.3 & 0.1 & 1 
& 0.8+0.2 & 0.1 &0 &0.1+0.0  & 0.0 & 
{\scriptsize HORIZONTAL} \\
\cline{2-10}
{\scriptsize Showers}& 8 & (1)+1.1+1.4 & 0.3 
& 2 & 0.2+0.8 & 0.2 &1&0.0+0.2&0.1& 
{\tiny ($-20^\circ\!\!<\delta<\!\!+20^\circ$)} \\
\cline{2-10}
\multicolumn{11}{c}{}\\[-2ex]
\cline{2-10}
{\scriptsize Tracks } & 7 &(6.4)+0.0+0.0 & 0.1 & 
0 &0.0+0.0 & 0.1  & 1  &0.0+0.0 & 0.0 &  
{\scriptsize DOWNGOING }\\
\cline{2-10}
{\scriptsize Showers}& 9 &(2.2)+0.1+0.6& 0.4 & 9 & 0.0+0.1
& 0.3 & 3&0.0+0.0&0.1&  
{\tiny ($-90^\circ\!\!<\delta<\!\!-20^\circ$)} \\
\cline{2-10}
\multicolumn{11}{c}{}\\[-2ex]
\cline{2-10}
{\bf Total}  & 35 &(12.6)+4.9+3.4& 1.1 
& 14 & 1.2+1.7 & 0.9 & 5  & 0.1+0.3 & 0.3 &  \\
\cline{2-10}
\multicolumn{11}{c}{}\\[-2.5ex]
 \multicolumn{1}{c}{}
&\multicolumn{3}{c}{30--100~TeV} 
&\multicolumn{3}{c}{100--300~TeV} & \multicolumn{3}{c}{300--3000~TeV}
\end{tabular}
\end{center}
\vspace{-0.5cm}
\caption{Number of events at IceCube (1,347 days) implied by the atmospheric
neutrino flux (conventional plus charm) and by the diffuse galactic
neutrino flux. We have assumed a pure He composition in the
CR flux at $E>E_{\rm knee}$. Events in parentheses correspond to
atmospheric muons entering the detector from outside. 
\label{tab1}}
\end{table}
Our estimate includes the attenuation by the Earth of the 
neutrino flux reaching IceCube from different zenith angles, 
the (energy and flavor-dependent)
effective volume of the detector \cite{Aartsen:2013jdh}, 
and the veto due to the 
accompanying muon in downgoing atmospheric events \cite{Aartsen:2014muf}.
Events in parentheses are
not genuine neutrino interactions 
but atmospheric muons entering the detector from outside. 
The uncertainty in the atmospheric background estimated by
IceCube \cite{Aartsen:2015zva}
for these three energy bins is around 8.6, 1.1 and 0.1 events,
respectively.

The comparison with the data 
reveals that there is an excess that is {\it (i)} more significant 
at higher energies and {\it (ii)} stronger
for showers than for tracks and for
down-going and near horizontal directions than for upgoing events
(see \cite{Palladino:2016xsy} for a fit to the anisotropy between 
the North and the South skies). 
For example, there are
a total of 14 events with a deposited energy between 100 and 300 TeV
but just $2.9\pm 1.1$ expected from atmospheric neutrinos.
The table also reflects that galactic neutrinos only provide around 2 
events among the 54 data sample. Our calculation of the
galactic signal 
seems robust, as we reproduce within a 10\% the signal estimated
by IceCube for different astrophysical neutrino 
fluxes (their Fig.~3, page 49 in \cite{Aartsen:2015zva}). 
Although our result is significantly larger than the one obtained
by other authors\footnote{This discrepancy is mostly due to the calculation
of IceCube events implied by a given neutrino flux, {\it i.e.}, to the 
cuts and the effective volume of the detector at different energies 
and event topologies (shower or track).} 
({\it e.g.}, only 0.1 galactic events among IceCube's three
year data in \cite{Joshi:2013aua}), it is still unable to explain
the high-energy IceCube excess. It is also consistent with a
recent likelihood analysis \cite{Denton:2017csz}
 indicating that the galactic contribution to
total IceCube signal is subleading (around $7\%$) or with
the total estimate (including point-like sources) in \cite{Ahlers:2015moa}
(4--8\%).

\section{Other components in the neutrino flux}
Let us go back to the simple scheme for the 
galactic CR flux outlined in Section 1,
with a spectral index
$\alpha^{A}=\alpha^{A}_0+\delta$ for each species $A$.
It is important to notice that, up to collisions and 
energy loss, the effects of the 
propagation will be identical for CRs with the same 
rigidity (same value of $E/Z$). A He 
nucleus ($Z=2$) of energy $E$ will describe exactly the
same trajectory through the galaxy as a proton of
energy $E/2$. This implies a relationship between the 
fluxes $\Phi_{p,{\rm He}}(E)$ that we see at the Earth 
and the production rate $I_{p,{\rm He}}(E)$ of each species 
at the sources:
\beqa
\Phi_p & = & 
\left(n\, \left(E/1\right)^{-0.5}\right)\times 
I_{p}\nonumber \\
\Phi_{\rm He} &=& 
\left(n\, \left(E/2\right)^{-0.5}\right)\times 
I_{\rm He}\,,
\label{prop}
\eeqa
where the factor multiplying $I_{A}(E)$ includes the overall
normalization $n$ and the propagation effects for $\delta=0.5$. 
Therefore, taking the fluxes in Eqs.~(\ref{fluxp},\ref{fluxHe}) 
we obtain that at $E<E_{\rm knee}$
the relative production rate of CRs by our galaxy is
\beqa
I_p & = & C \;
E^{-2.2}\nonumber \\
I_{\rm He} &=& 0.29\,C\;
E^{-2.1}\,.
\label{feg1}
\eeqa
Notice that within this scheme CRs would 
stay confined in the galaxy for a period of order 
\beq
\tau_G\approx 10^{7}\;{\rm years} \times \left( {E \over 
Z\times 10\;{\rm GeV}} \right)^{-0.5}\,.
\eeq
We then assume  a steady state, {\it i.e.}, 
the number of CRs leaving the galaxy 
is similar to the number of CRs accelerated at the 
sources. This means that 
at energies below $E_{\rm knee}$ our galaxy is emitting protons
and He nuclei at the rate given in Eq.~(\ref{feg1}). 
Once emitted, these 
CRs will stay inside the cluster and supercluster 
--generically, the intergalactic (IG) medium--
containing our galaxy for a time that may be 
larger than the age of 
the universe \cite{Berezinsky:1996wx,Harari:2016vtz}.

A similar argument applied to the CR flux at $E>E_{\rm knee}$
in Eq.~(\ref{fluxA}) implies a galactic production/emission rate
\beq
I_A  = {254\over \sqrt{Z}}\,C \;
E^{-2.5}\,,
\label{feg2}
\eeq
where $Z$ is the corresponding atomic number of the CRs 
dominating the galactic flux at these energies.
Our basic statement in Eqs.~(\ref{feg1},\ref{feg2}) is that the
spectral index and the relative composition of the CRs emitted
by our galaxy into the IG medium are correlated with the ones 
we see reaching the Earth, and that this
correlation depends on a single transport parameter 
$\delta$ that, in the simple scenario under consideration, 
takes the value $\delta=0.5$. 

Let us now assume that galaxies 
(the supernova remnants and pulsars inside them)
are the main source of CRs of
energy up to $10^8$ GeV, and that 
ours is an {\it average} galaxy. CRs can then be found {\it (i)} inside
the galaxies (including ours), where they exhibit 
the spectrum and composition 
in Eqs.~(\ref{fluxp}--\ref{fluxA}), and {\it (ii)} in the 
IG space, where they appear with 
the spectrum and the composition in
Eqs.~(\ref{feg1},\ref{feg2}). The interactions 
of these two types of CRs with the gas in the medium
where they propagate will produce 
TeV--PeV neutrinos. Therefore, 
in the astrophysical neutrino flux discovered by 
IceCube we may consider the relative weight of the following
three components:

\begin{itemize}

\item Neutrinos from CR interactions with the IS matter
in our own galaxy. This component, $\Phi_\nu^{\rm gal}$, has been
discussed in the previous section, and it is 
way too low to account for the number of events detected at IceCube.
In addition, these neutrinos are 
concentrated near the galactic plane.

\item Neutrinos from the same type of interactions but
in other galaxies. 
As discussed above, 
if  both the accelerators and
the spectrum of magnetic turbulences 
are universal we may expect that the IS medium in other galaxies
will confine CRs with the same spectrum and relative composition 
as in ours, given in Eqs.~(\ref{fluxp}--\ref{fluxA}).  
Collisions 
with the  gas there will then produce a neutrino flux from 
all galaxies, 
$\Phi_\nu^{\rm AG}$, proportional to the one in 
Eqs.~(\ref{fluxnuG0},\ref{fluxnuG1}). 
Such flux will be more
isotropic than the one discussed in the previous section, but 
its $\approx E^{-2.6}$ spectrum seems too steep to account
for a significant fraction of the IceCube events. 
In particular, it has been shown
\cite{Murase:2013rfa} (see also \cite{Stecker:1978ah}) that  
the neutrino flux 
would come together with a 10--100 GeV diffuse gamma-ray flux 
inconsistent with 
Fermi-LAT data \cite{Abdo:2010nz}.

\item Neutrinos from interactions of CRs with 
extragalactic gas \cite{Berezinsky:1996wx,Harari:2016vtz,
DeMarco:2005va,Murase:2012rd,Zandanel:2014pva}. 
As mentioned before, the CRs producing this IG
neutrino flux $\Phi_\nu^{\rm IG}$
are steadily emitted by all  the 
galaxies with the spectrum and composition 
in Eqs.~(\ref{feg1},\ref{feg2}). 
In the intracluster space these CRs will
face a gas density typically $10^{-4}$ times smaller than 
the one inside the parent galaxy, but the time they spend
there may be $10^{-5}$ times larger, resulting into a
larger column density. In addition,
while inside galaxies CRs are in a steady state, in the IG
medium the total number of CRs grows with time.

\end{itemize}

In the next sections we will calculate $\Phi_\nu^{\rm IG}$ 
up to an overall normalization factor and will show that their
spectrum may provide a good fit of the high-energy IceCube
data, including the non-observation of the Glashow resonance
at 6.3 PeV. 

\section{Diffuse flux of intergalactic neutrinos}
Our starting point is a CR number density
in IG space  with the 
spectrum and the relative composition given by 
Eqs.~(\ref{feg1},\ref{feg2}). Notice that an isotropic flux $\Phi_A$ of 
(relativistic) CRs type $A$ would be simply related to
the number density $n_A$ by $\Phi_A = n_A\;c/(4\pi)$.
Let us assume that the average gas density  
in the IG medium is $\bar \rho_{\rm IG}$, with a $(1\!:\!3)$ He to H ratio.
The neutrino flux reaching the Earth from CR collisions 
with the gas along the line of sight is then \cite{Carceller:2016upo} 
\beq
\bar \Phi_\nu(E) = R\, \bar \rho_{\rm IG} \,\sum_A \frac{F_A}{m_p} 
\int^1_0 {\rm d}x \, \sigma_{A p}(E/ x)\, 
\bar \Phi^{\rm IG}_A(E/x)\, x^{-1} f^{\nu}_{A}(x,E/x) \,.
\label{final1}
\eeq
where $R$ is the maximum distance\footnote{We neglect the 
redshift in the contribution to the neutrino flux from
distant clusters.} in our supercluster and 
beyond, $A$ runs over the different species
in the CR flux, $f^{\nu}_{A p}(x,E')$ is the yield of neutrinos
carrying a fraction $x$ of the incident energy produced in 
$A\,p$ collisions, and $F_A$ takes into account the mixed
H/He composition of the IG gas
($F_p\approx 0.92$, $F_{\rm He}\approx 0.90$ and 
$F_{\rm Fe}\approx 0.86$ \cite{Carceller:2016upo}).
This expression gets simplified if one neglects the 
energy dependence of the yields and takes an unbroken
power law both for the intergalactic CR flux
[$\bar \Phi^{\rm IG}_{A}(E)=n\, I_A= n_A E^{-\alpha_A}$, with $I_A$ 
given in Eqs.~(\ref{feg1},\ref{feg2})] and for the cross 
section [$\sigma_{A p}(E) = \sigma^0_{A p} E^{\beta_{A}}$]:
\beq
\bar \Phi_\nu^{\rm IG}(E) = R\, \bar \rho_{\rm IG} 
\sum_A \frac{F_A \, \sigma^0_{A p}\, n_A }{m_p} \;
 Z^{\nu}_{A}\, E^{-(\alpha_A-\beta_{A})} \,,
\label{final0}
\eeq
being $Z^{\nu}_{A}$ the order-$(\alpha_A-\beta_{A}-1)$  moment
of the yield,
\beq
Z^{\nu}_{A}=
\int^1_0 {\rm d}x\, x^{\alpha_A-\beta_{A}-1} \,f^{\nu}_{A}(x) \,.
\label{zmoment}
\eeq
\begin{figure}[!t]
\begin{center}
\includegraphics[width=0.48\linewidth]{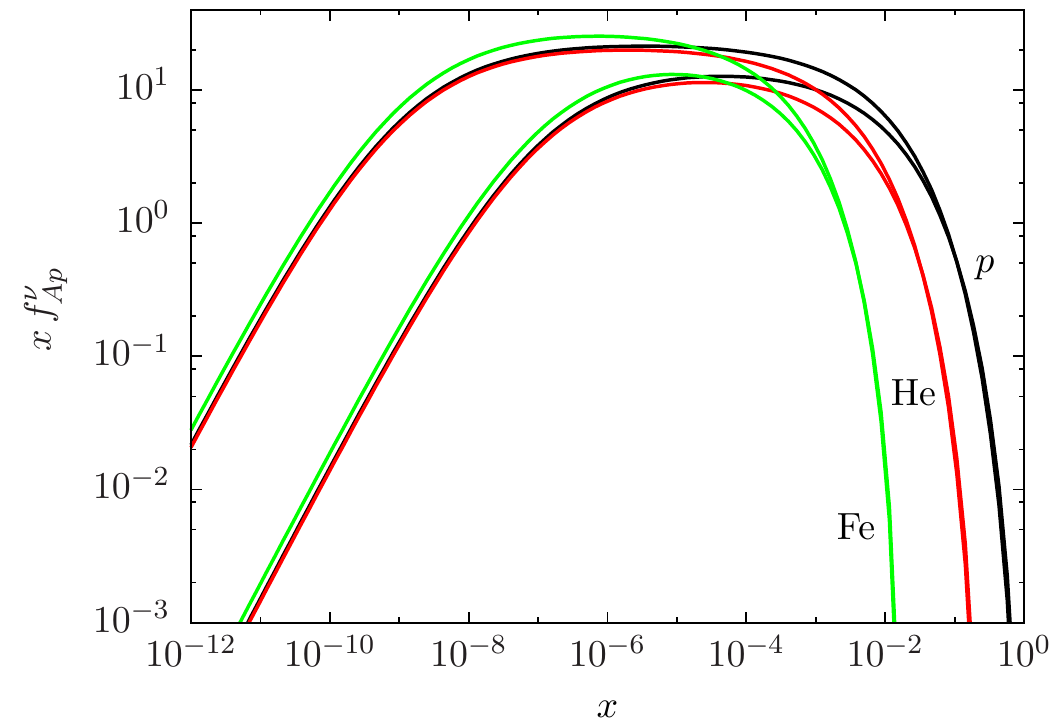}
\end{center}
\caption{Total neutrino yield ($\nu$ and $\bar \nu$ of all flavors)
$f^{\nu}_{A}(x,E)$ from
$A=$ proton, helium and iron collisions
with a proton at rest at $E=10^{6},10^{8}$ GeV \cite{Carceller:2016upo}.
\label{f2}}
\end{figure}
We see that the energy dependence of the cross sections 
will slightly change the spectral index of the IG 
neutrino flux from  
$\alpha_A$ to $\alpha_A-\beta_A$. 
In $p\,p$ collisions we have $\beta_p=0.082$ and 
$\sigma^0_{p p}=17.7$ mb, whereas in 
He$\,p$ and Fe$\,p$ collisions 
$\beta_{\rm He}=0.062$, $\sigma^0_{{\rm He}\, p}=60.5$ mb, 
$\beta_{\rm Fe}=0.026$ and $\sigma^0_{{\rm Fe}\, p}=551$ mb.
Taking the yields from \cite{Carceller:2016upo} and
encapsulating the unknowns in a single normalization
factor $N$, at $E<10^{5.5}$ GeV we obtain 
\beq
\bar \Phi^{\rm IG}_\nu = 
2.8 \,N\, E^{-2.12}
+
1.0 \,N\, E^{-2.04}
\,,
\label{fluxnuIG0}
\eeq
where the two terms come from the proton
and the He contributions, respectively.
At neutrino energies $E>10^{6.5}$ GeV we find
\beq
\bar \Phi^{\rm IG}_\nu = \left\{\begin{array}{ll}
290 \,N\, E^{-2.42}
& {\rm \; (100\%\;proton)}\,,\\ 
106 \,N\, E^{-2.44}
& {\rm \; (100\%\;helium)}\,,\\ 
11 \,N\, E^{-2.47}
& {\rm \; (100\%\;iron)}\,.
\end{array}\right.\,
\eeq
The large uncertainty in $\Phi^{\rm IG}_\nu$ is related to the
CR composition: its origin is the $Z$-moment of the
neutrino yield, which is much smaller for heavy nuclei 
than for protons (see Fig.~\ref{f2}). If primary
CRs above $E_{\rm knee}$ were mostly protons, then 
$\Phi^{\rm IG}_\nu$ would be 
3.8 times larger than if they are pure helium, but this
neutrino flux could
also be a factor of 0.056 smaller if CRs were 100\% iron.

In order to simplify our analysis, beyond $E_{\rm knee}$
we will consider the IG  flux 
\beq
\bar \Phi^{\rm IG}_\nu = a\, N\, E^{-2.44}\,
\label{fluxnuIG1}
\eeq
with $6<a<400$, and we will use a 
power law to interpolate 
between this flux and the one in Eq.~(\ref{fluxnuIG0}) at 
$E<10^{5.5}$ GeV. Notice that the same
value of $a$ may result from different CR compositions; for example,
$a=106$ could correspond to 100\% He or to 
$25\%$ proton plus $75\%$ iron.
In the next section we will fit the high-energy IceCube data
with the parameters $N$ and $a$ in this
neutrino flux.

\section{Fit of the high-energy IceCube data}
Let us take the average IG flux obtained in the previous section 
to be isotropic.\footnote{The flux could actually be modulated 
by the large-scale structure around our 
galaxy.}
In the first row of Table~\ref{tab2} we write the IceCube excess in each 
energy bin, {\it i.e.}, the difference between the 
data and the sum of the atmospheric and the galactic events given 
in Table~\ref{tab1}, including IceCube's estimate of the background uncertainty.
In the second row
we give the number of events predicted by an unbroken power law
with spectral indices $2.0$, which was the initial 
neutrino flux proposed by IceCube after three years of observations.
In the third and fourth rows of Table~\ref{tab2}
we give the number of events predicted by the IG flux 
$\Phi_\nu^{\rm IG}$ with $a=106$ and by
the all-galaxies flux $\Phi_\nu^{\rm AG}$, which basically consists
of the galactic flux in Eqs.~(\ref{fluxnuG0},\ref{fluxnuG1}) but
isotropic and with an arbitrary normalization. 

\begin{table}[!t]
\begin{center}
\begin{tabular}{r|c|c|c|c|c|}
          \multicolumn{1}{c}{}
        & \multicolumn{1}{c}{ 30--100~TeV } 
        & \multicolumn{1}{c}{ 0.1--0.3~PeV }
        & \multicolumn{1}{c}{ 0.3--3~PeV } 
        & \multicolumn{1}{c}{ 3--10~PeV } 
\\
\multicolumn{5}{c}{}\\ [-3ex]
\cline{2-5} 
{ Excess } & 13.0$\pm$8.6 & 10.2$\pm$1.1 & 4.3$\pm$0.1 & 0 \\ [0ex]
\cline{2-5}
\multicolumn{5}{c}{}\\ [-2ex]
\cline{2-5}
{ $E^{-2.0}$ } & 4.7 & 6.4 & 8.1 & 3.2 \\
\cline{2-5}
{ IG } & 7.6 & 9.3 & 5.2 & 0.4  \\
\cline{2-5}
{ AG } & 14.0 & 11.0 & 3.5 & 0.1 \\
\cline{2-5}
\end{tabular}
\end{center}
\vspace{-0.5cm}
\caption{Total number of events for 1,347 days 
at IceCube implied by the 
fluxes discussed in the text together with the excess 
deduced from Table~\ref{tab1}.
\label{tab2}}
\end{table}
The normalization of each astrophysical flux has 
been fixed so that they reproduce the total IceCube excess at 
100 TeV--3 PeV ({\it i.e.}, the sum of the two high energy bins
in Table~\ref{tab1}).
The flux $\Phi_\nu^{\rm AG}$ would imply a too strong 
gamma-ray signal \cite{Stecker:1978ah}. In fact, as shown in \cite{Murase:2013rfa},
any neutrino flux steeper than $E^{-2.2}$ would appear with a gamma-ray
flux that extrapolated to lower energies conflicts the Fermi-LAT data
\cite{Abdo:2010nz}. To be realistic those fluxes would require a mechanism 
that absorbs the gammas
leaving the neutrino flux unaffected \cite{Murase:2015xka}. 
Another interesting possibility for the AG flux 
could be a radially dependent $\delta$ parameter \cite{Gaggero:2015xza}
that would make the harder the CR flux near the galactic
center, where most of the interactions occur.

Notice that in Table~\ref{tab2} we have added a fourth energy bin, 
3--10 PeV, which provides an important piece of information
in order to decide about the goodness of the fits. At these energies
electron  antineutrinos could reveal the 
Glashow resonance through collisions with electrons:
\beq
\bar \nu_e \, e \to W^- \to q \, \bar q\,,\; \ell \, \bar \nu_\ell
\eeq
\begin{figure}[!t]
\begin{center}
\includegraphics[width=0.5\linewidth]{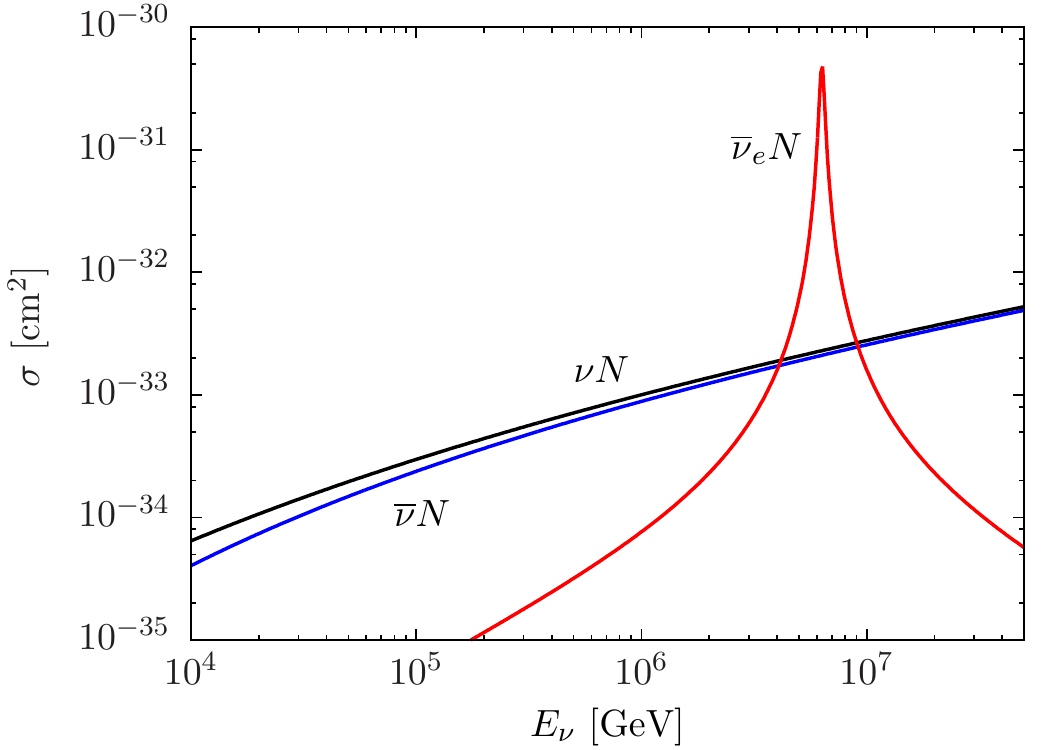}
\end{center}
\caption{Total $\nu\, N$, $\bar \nu\, N$ and $\bar\nu_e\, e$
cross sections.
\label{f3}}
\end{figure}
In Fig.~\ref{f3} we show that at $E=(6.3\pm 2.0)$ PeV
the cross section for this process \cite{Bhattacharya:2011qu} goes 
well above $\sigma(\nu N)$ \cite{Connolly:2011vc}. 
Since the IceCube
target has 10 electrons per 18 nucleons and the $\bar\nu_e$
frequency in the IG neutrino flux is almost exactly 1:6 
[see Eq.~(\ref{flav2})], the Glashow resonance
will clearly have an impact on the fit. Notice also that when the
$W$ decays hadronically 
(with a 67.6\% branching ratio)
all the neutrino energy $E_\nu$ will be deposited in the ice, while in 
leptonic decays (32.4\% of the times) the charged lepton will take 
an energy between 0 and $E_\nu$ with an average value of 
$0.33\,E_\nu$ \cite{Bhattacharya:2011qu}. 

We find that the $E^{-2.0}$ flux implies 3.2 events 
beyond 3 PeV, while all the other fluxes may fit the
data while predicting less than one event in that bin. 
Therefore, the non-observation of
the Glashow resonance after four years of data 
disfavors the harder $E^{-2.0}$ flux initially 
proposed by IceCube. The physically motivated flux 
$\Phi_\nu^{\rm IG}$ has a similar spectral index at 
lower energies [see Eq.~(\ref{fluxnuIG0})], however,
a possible break caused by a change
in the CR composition at $E_{\rm knee}$ may define an acceptable
possibility. Indeed, the presence of the knee in the CR spectrum implies
that we should {\it not} expect an unbroken power law
in the neutrino flux at TeV--PeV energies. Although the amount of IceCube
data is small, a recent analysis \cite{Anchordoqui:2016ewn}
that includes bounds from
Fermi-LAT data excludes at the $3\sigma$ level an astrophysical neutrino
flux described by a single power law,
favoring a break in the spectrum at 200--500 TeV very similar to
the one we obtain here.

\begin{figure}[!t]
\begin{center}
\includegraphics[width=0.48\linewidth]{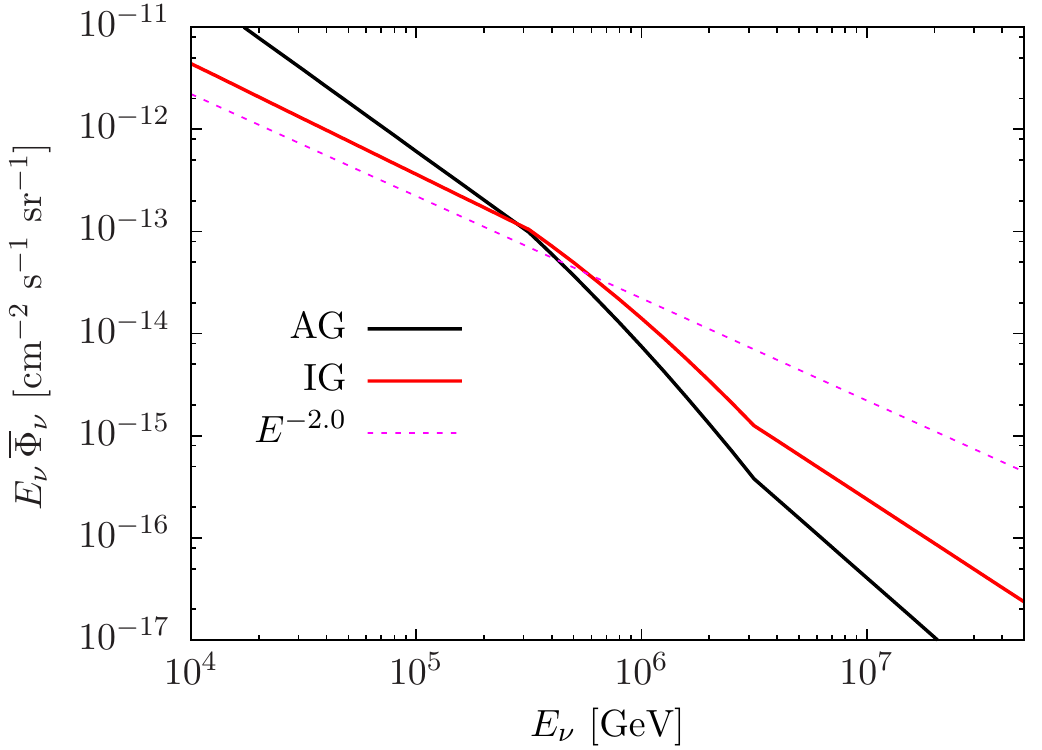}\hspace{0.6cm}
\includegraphics[width=0.46\linewidth]{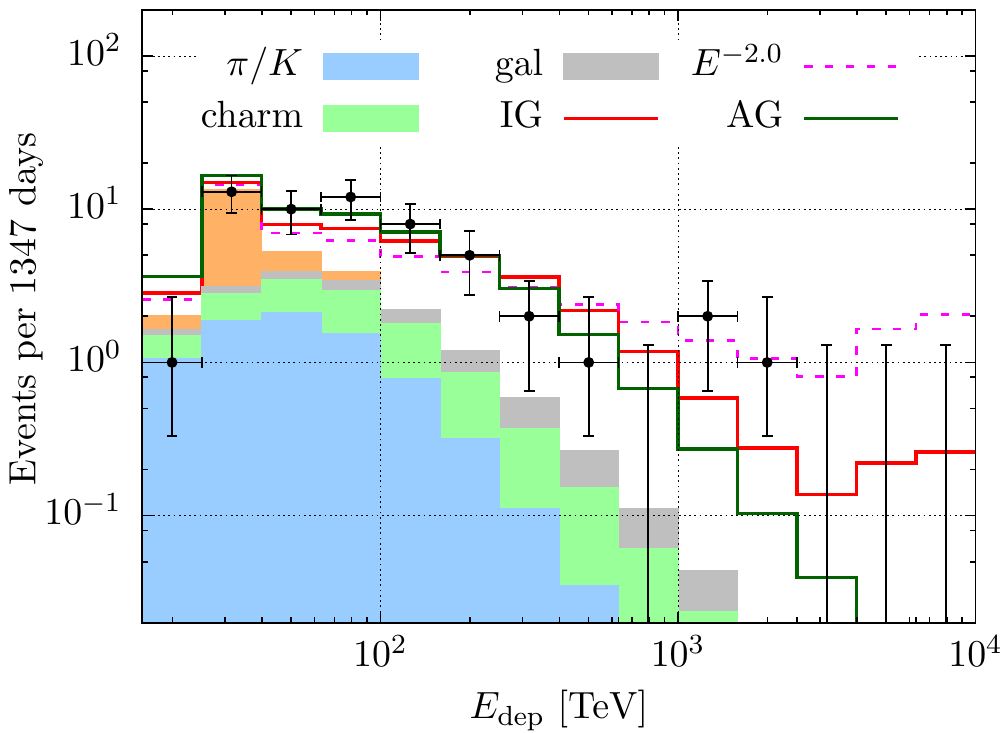}
\end{center}
\caption{{\bf Left.} Different components in the neutrino flux reaching
the Earth. The IG and AG fluxes correspond to a 
dominant helium composition in the CR flux at $E>E_{\rm knee}$ (see text).
{\bf Right.} Event distribution at IceCube (1,347 days) for the three diffuse
fluxes (IG, AG and $E^{-2.0}$) including the 
background of atmospheric muons entering the detector from
outside (orange contribution).
\label{f4}}
\end{figure}

In Fig.~\ref{f4} we plot these astrophysical
$\nu$ fluxes together
with the total number of events that they imply at all
IceCube energies.
For comparison, the normalization of the atmospheric 
and galactic neutrinos fluxes in Fig.~\ref{f1} reads
\beqa
\bar \Phi_\nu^{\pi/K} (100\;{\rm TeV}) &=& 5.1\times 10^{-18} \;
{\rm (GeV\,cm^2\,sr\,s)^{-1}}\,,\nonumber \\
\bar \Phi_\nu^{{\rm charm}} (100\;{\rm TeV}) &=&  1.9\times 10^{-18} \;
{\rm (GeV\,cm^2\,sr\,s)^{-1}}\,,\nonumber \\
\bar \Phi_\nu^{\rm gal} (100\;{\rm TeV}) &=&  4.9\times 10^{-19} \;
{\rm (GeV\,cm^2\,sr\,s)^{-1}}\,.
\eeqa
whereas the four fluxes in Fig.~\ref{f4} have
been normalized to
\beqa
\bar \Phi_\nu^{(2.0)} (100\;{\rm TeV}) &=&  2.2\times 10^{-18}\; 
{\rm (GeV\,cm^2\,sr\,s)^{-1}}\,,\nonumber \\
\bar \Phi_\nu^{(2.58)} (100\;{\rm TeV}) &=&  5.2\times 10^{-18} \;
{\rm (GeV\,cm^2\,sr\,s)^{-1}}\,,\nonumber \\
\bar \Phi_\nu^{\rm IG} (100\;{\rm TeV}) &=&  3.6\times 10^{-18} \;
{\rm (GeV\,cm^2\,sr\,s)^{-1}}\,,\nonumber \\
\bar \Phi_\nu^{\rm AG} (100\;{\rm TeV}) &=&  6.1\times 10^{-18} \;
{\rm (GeV\,cm^2\,sr\,s)^{-1}}\,.
\eeqa

A final comment concerns the possible North--South sky asymmetry of the IceCube
signal in Table 2. Let us focus on the three bins of energy  above 100 TeV,
where the uncertainties are lower. The 14.5 excess in Table 2 is distributed as follows:
1.0 events from upgoing directions (Northern sky, with declinations 
$90^\circ > \delta > 20^\circ$), 1.3 from near-horizontal directions
($20^\circ > \delta > -20^\circ$), and 12.2 from downgoing directions
(Southern sky, $-20^\circ > \delta > -90^\circ$). In the same direction
bins, the (isotropic) IG flux implies 2.1, 5.1, and 7.2 events, respectively,
whereas the (also isotropic) AG flux would give 2.4, 5.9 and 6.2 events.
Therefore, as emphasized in \cite{Palladino:2016xsy}, although the 
sample is still small, the data seems
to favor an anisotropic neutrino flux. For an IG origen this
could simply reflect a larger total column density of 
intracluster gas along the directions in the Southern sky.

\section{Dependence on the cosmic ray composition}
The distribution of the IceCube data given 
in Fig.~\ref{f4} shows a deviation from 
a power law at energies above 250 TeV. In particular, in the five 
bins between $10^{2.4}$ and $10^{3.4}$ TeV we find, respectively,
2, 1, 0, 2 and 1 events. This sequence suggests a flat event distribution following
the steeper one observed at lower energies, or even a possible
drop in the event rate at 0.25--1 PeV followed by a
{\it bump} defined by the three events of highest energy.
Obviously, the statistical significance of such deviations is
limited (notice that in Tables 1 and 2 we have defined much wider
bins in order to {\it dilute} them), but it is apparent that none of
the neutrino fluxes that we have discussed so far
would be able to accommodate such a spectral feature.
Remarkably, the  new IceCube analysis presented in ICRC2017
\cite{ICRC2017} corresponding to 
two more years of data taking does not include new events of $E>250$ TeV
among a 28 event sample, which tends to steepen the neutrino
flux and/or to increase the significance of the 0.25-1 PeV drop.

Therefore, it may be
interesting to study what type of spectral irregularities may be expected
from sudden changes in the CR composition, which could be 
associated, for example, 
to the maximum energy achieved by 
cosmic accelerators for CRs of a given charge. We will see that
these spectral changes in the secondary 
neutrino flux would appear even when the
primary CR flux exhibits an unbroken power law. 

\begin{figure}[!t]
\begin{center}
\includegraphics[width=0.5\linewidth]{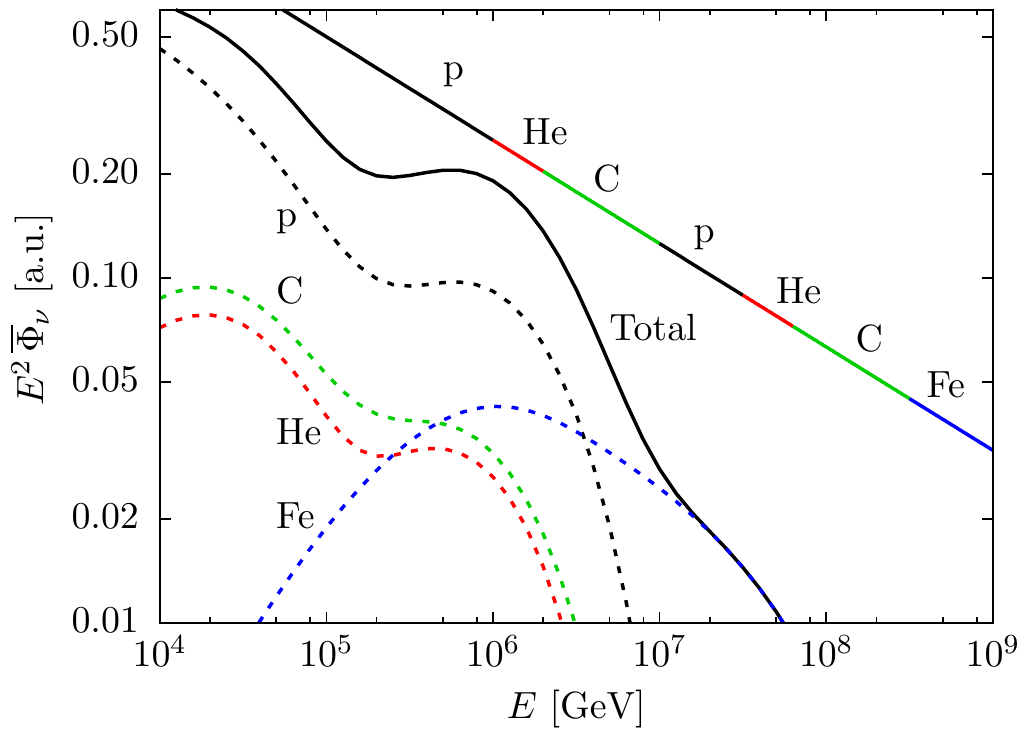}
\end{center}
\caption{Neutrino flux (in arbitrary units) implied by a CR flux
of spectral index 2.3 and changing composition. We plot the contribution of
each species and the total neutrino flux.
\label{f5}}
\end{figure}

To be definite, let us consider a CR flux proportional to $E^{-2.3}$ in the whole
$10^4$--$10^{10}$ GeV energy interval. We will assume (see Fig.~\ref{f5})
that the dominant composition
is proton up to $10^6$ GeV, then He up to $2\times 10^6$, and C up to $10^7$ GeV.
At this energy the flux becomes proton dominated again up to $10^{7.5}$ GeV, then
He up to $2\times 10^{7.5}$ GeV, C up to $10^{8.5}$ GeV, and Fe at higher energies.
As expressed in Eq.~(\ref{final1}), the secondary neutrino flux will then depend on 
the fraction of each species in this CR flux that interacts   
with IG matter (this fraction is proportional to the 
cross section) and on the neutrino yield
in those interactions. In  Fig.~\ref{f5} we plot the neutrino
flux up to an overall normalization factor. For comparison, we include in the plot
the parent CR flux (the relative normalization between both fluxes is also arbitrary).

The key observation is that 
protons and nuclei of energy $E$ contribute to neutrinos of energy 
below $0.1E$ and $0.1A^{-1}E$, respectively. As a consequence,
a change in the composition 
towards heavier nuclei tends to produce a drop (relative to a constant
spectral index) in the neutrino flux, whereas a change in the opposite direction 
--heavy to light-- may introduce a bump. In the plot we see
that the change from proton to He and then to carbon at $E>1$ PeV translates into
a neutrino drop in the energy region 
suggested by the data, whereas the change from C to p at 
$10$ PeV could induce a relative excess at energies around 1 PeV.
The subsequent changes to heavier nuclei at $E>3$ PeV would be motivated 
by the absence in the IceCube data of the 6.3 PeV Glashow resonance.
Although this CR flux is just a toy model,  it shows that 
the CR composition at energies around $E_{\rm knee}$ and
beyond may be a key factor to justify deviations from a power law in
the high-energy neutrino flux detected by IceCube.

\section{Summary and discussion}
High energy neutrinos can only be produced in the collisions
of charged CRs. It seems then clear that 
the discovery at IceCube of an astrophysical neutrino flux 
will have implications in our understanding of high energy CRs. 
In particular, a higher statistics should 
establish the spectral index of this flux at $E\lsim 1$ PeV
and, most important, the presence or not of the Glashow 
resonance at $E\approx 6.3$ PeV.
We have shown that these two observations will provide clear hints
about the spectrum and the composition of the parent CRs,
which in turn relate to the environment where the neutrinos
have been produced.

Galactic CRs are described by a spectrum $\approx E^{-\alpha}$ 
that is steeper than the one they have at the sources: 
$\alpha_0=\alpha-\delta$ with $\delta=0.5$ in the simplest
scenario. 
The neutrinos produced in their collisions will inherit 
the spectral index of the parent CR flux. If
the main source of the IceCube neutrinos were the collisions of 
CRs {\it inside} galaxies, then their spectral index  
would be $2.6$ at 
$E<10^{5.5}$ GeV and around $2.9$ at higher energies.
A few 1--2 PeV events at IceCube from such steep 
flux would then be correlated
with a too large diffuse gamma-ray flux at 0.1--100 GeV 
\cite{Stecker:1978ah,Murase:2013rfa}.
Once CRs leave into the IG space, however, 
their spectral index should be similar to the one they
have at the sources. In Eq.~(\ref{fluxnuIG0}) we provide a two-component
(from proton and He collisions) IG neutrino flux with a spectral 
index near $2.1$. Such a hard spectrum, if unbroken, should have already
revealed the Glashow resonance.
We have shown, however, that if the CR knee brings a change in the 
composition towards heavier nuclei then the secondary neutrino flux 
may experience a sudden 
drop at $E>1$ PeV. Therefore, the 
observation (or not) of the Glashow resonance will provide
important information about the CR composition at these
energies.

As a viable possibility, we have studied the 
implications at IceCube of the IG
neutrino flux that may appear if CRs above $E_{\rm knee}$ are dominated by He 
($a=106$ in Eq.~(\ref{fluxnuIG1})). Our results are summarized
in Table~\ref{tab2}. We see that,
normalizing the neutrino flux so that the total number of
events in the two high energy bins matches the experimental
excess, this single component $\Phi_\nu^{\rm IG}$
provides a good fit of all the data. We find that the
(much steeper) galactic diffuse flux $\Phi_\nu^{\rm gal}$ contributes 
with just two events
in the IceCube sample, a number that is significantly 
larger than previous
estimates by other authors. 
Within our scheme, a pure proton composition above 
the CR knee is disfavored as $\Phi_\nu^{\rm IG}$ would imply around
2 events of $E>3$ PeV. 

Our analysis depends basically on the transport 
parameter $\delta$. The value $\delta=0.5$ that we have considered
is consistent with 
a Kraichnan spectrum of magnetic turbulences and diffusive 
shock acceleration at supernova remnants, although other
possibilities could be accommodated. We have also discussed 
possible deviations from a power law in the neutrino flux caused
by sudden changes in the primary CR composition.
Therefore, we think that
the astrophysical neutrino flux discovered by IceCube, once it is 
fully characterized, will provide very valuable information that will
help to complete the CR puzzle.

\section*{Acknowledgments}
We would like to thank Eduardo Battaner for discussions.
This work has been supported by MICINN of Spain 
(FPA2013-47836, FPA2015-68783-REDT, FPA2016-78220 and  
Consolider-Ingenio {\bf MultiDark} CSD2009-00064) and by Junta de 
Andaluc\'\i a (FQM101). JMC acknowledges a {\it Beca de
Iniciaci\'on a la Investigaci\'on} fellowship from the UGR.

\end{document}